\documentclass[aps,prl,preprint,superscriptaddress]{revtex4-1}
\usepackage{url}
\usepackage{amsmath}
\usepackage{lineno}
\usepackage{physics}
\usepackage{graphicx}
\usepackage{lmodern} %fix some errors about font sizes

\newcommand{\SIO}{Sr$_2$IrO$_4$}

\begin{document}

% Page header
%\markboth{Author et al.}{Short title}

% Title
\title{Square Lattice Iridates}

%Authors, affiliations address.
\author{Joel Bertinshaw}
\affiliation{Max Planck Institute for Solid State Research, 
Heisenbergstra\ss e 1, D-70569 Stuttgart, Germany}
\author{Y. K. Kim}
\affiliation{Graduate School of Nanoscience and Technology, Korea Advanced Institute of Science and Technology (KAIST), Daejeon 34141, South Korea}
\affiliation{Department of Physics, Korea Advanced Institute of Science and Technology (KAIST), Daejeon 34141, South Korea}
\author{Giniyat Khaliullin}
\affiliation{Max Planck Institute for Solid State Research, 
Heisenbergstra\ss e 1, D-70569 Stuttgart, Germany}
\author{B. J. Kim}
\email{bjkim6@postech.ac.kr}
\affiliation{Max Planck Institute for Solid State Research, 
Heisenbergstra\ss e 1, D-70569 Stuttgart, Germany}
\affiliation{Department of Physics, Pohang University of Science and Technology, Pohang 790-784, Republic of Korea}
\affiliation{Center for Artificial Low Dimensional Electronic Systems, Institute for Basic Science (IBS), 77 Cheongam-Ro, Pohang 790-784, Republic of Korea}

%Abstract
\begin{abstract}
Over the last few years, \SIO, a single-layer member of the Ruddlesden-Popper series iridates, has received much attention as a close analog of cuprate high-temperature superconductors. Although there is not yet firm evidence for superconductivity, a remarkable range of cuprate phenomenology has been reproduced in electron- and hole-doped iridates including pseudogaps, Fermi arcs, and $d$-wave gaps. Further, a number of symmetry breaking orders reminiscent of those decorating the cuprate phase diagram have been reported using various experimental probes. We discuss how the electronic structures of \SIO\ through strong spin-orbit coupling leads to the low-energy physics that had long been unique to cuprates, what the similarities and differences between cuprates and iridates are, and how these advance the field of high-temperature superconductivity by isolating essential ingredients of superconductivity from a rich array of phenomena that surround it. Finally, we comment on the prospect of finding a new high-temperature superconductor based on the iridate series.
\end{abstract}

%Keywords, etc.
%\begin{keywords}
%high-temperature superconductivity, spin-orbit coupling, %cuprates, iridates, pseudogap, intertwined orders
%\end{keywords}
\maketitle

%Table of Contents
\tableofcontents

%Section 1: Introduction
\section{INTRODUCTION}\label{sec-intro}
Since the discovery of the spin-orbit induced Mott state in 2008~\cite{Kim08,Kim09}, iridates have been at the center of an intensive search for novel phenomena that arise from the combined influence of spin-orbit coupling (SOC) and electron correlations~\cite{Pesi10,Witc14,Rau16,Scha16,Wint17,Herm18}. The new Mott phase has attracted much interest from both the topological insulator and the strongly correlated electron system communities. From the former perspective, the interactions open possibilities for novel topological phases of matter that arise in the presence of magnetic order, such as Weyl semimetals with surface Fermi arcs or Axion insulators with unusual electromagnetic responses. Here, pyrochlore iridates are prime candidates for the realization of novel topological phases~\cite{Wan11,Witc12}. In the context of correlated electron physics, SOC leads to anisotropic and highly frustrated spin interactions, which manifest themselves in extraordinary manners when they dominate over the isotropic interactions that usually dictate the properties of conventional magnets. A prominent example is the possible realization of the Kitaev spin liquid with fractional excitations and statistics. Honeycomb iridates have been extensively investigated for their relevance to the Kitaev model~\cite{Jack09,Choi12,Chal13,Hwan15,Chal15}, along with a recent extension of the materials scope to the $4d$ transition-metal compound RuCl$_3$~\cite{Plum14,Bane16}. The above two topics are discussed in recent reviews~\cite{Witc14,Rau16,Scha16,Herm18}.

The main topic of this review, \SIO, is distinct from the above threads in that it has a counterpart in weakly spin-orbit coupled systems, namely, cuprate high temperature superconductors (HTSC). Despite being central to the establishment of the spin-orbit Mott insulating phase, the effect of SOC is not very conspicuous in the physical properties of \SIO. The details of this rather exotic mechanism are reviewed in Section~2.

The essence of the cuprate physics is believed to be contained in its unique realization of a single-band spin-1/2 Heisenberg antiferromagnet (AF) on a quasi-two-dimensional (2D) lattice~\cite{Lee06}, with strong exchange couplings $J$$\sim$130 meV~\cite{Hayd91}. To the best of our knowledge, \SIO\ is the first material outside of the cuprate family that also realizes this condition (Fig.~1)~\cite{Kim12a}. This naturally leads to the expectation that \SIO\ can be a platform for unconventional superconductivity~\cite{Wang11,Wata13,Meng14,Yang14,Gao15}. Of course, no two materials share an identical electronic structure, and in fact the microscopic details of cuprates and \SIO\ are quite different in many aspects. 

To what extent does the analogy between cuprate and \SIO\ physics remain valid? Is \SIO\ as strongly correlated as cuprates, and are the spin dynamics indeed well described by spin-1/2 Heisenberg interactions? We address these issues in Section~2. Section~3 deals with more vital questions: does the analogy continue to hold away from the Mott insulating phase as charge carriers are introduced? Namely, is a single-band Fermi surface formed upon doping? Are there strong magnetic fluctuations dressing the quasiparticles and driving their mutual interactions?

While the above questions remain under investigation, experiments to date show that electron- and hole-doped square lattice iridates reproduce a remarkable range of phenomenology of cuprates. In Section~4, we review the nature of the pseudogap and $d$-wave gap observed by angle-resolved photoemission (ARPES) and scanning tunneling spectroscopy (STS), and evidence for symmetry breaking orders that possibly underlie the pseudogap.

\begin{figure}
\includegraphics[width=5in]{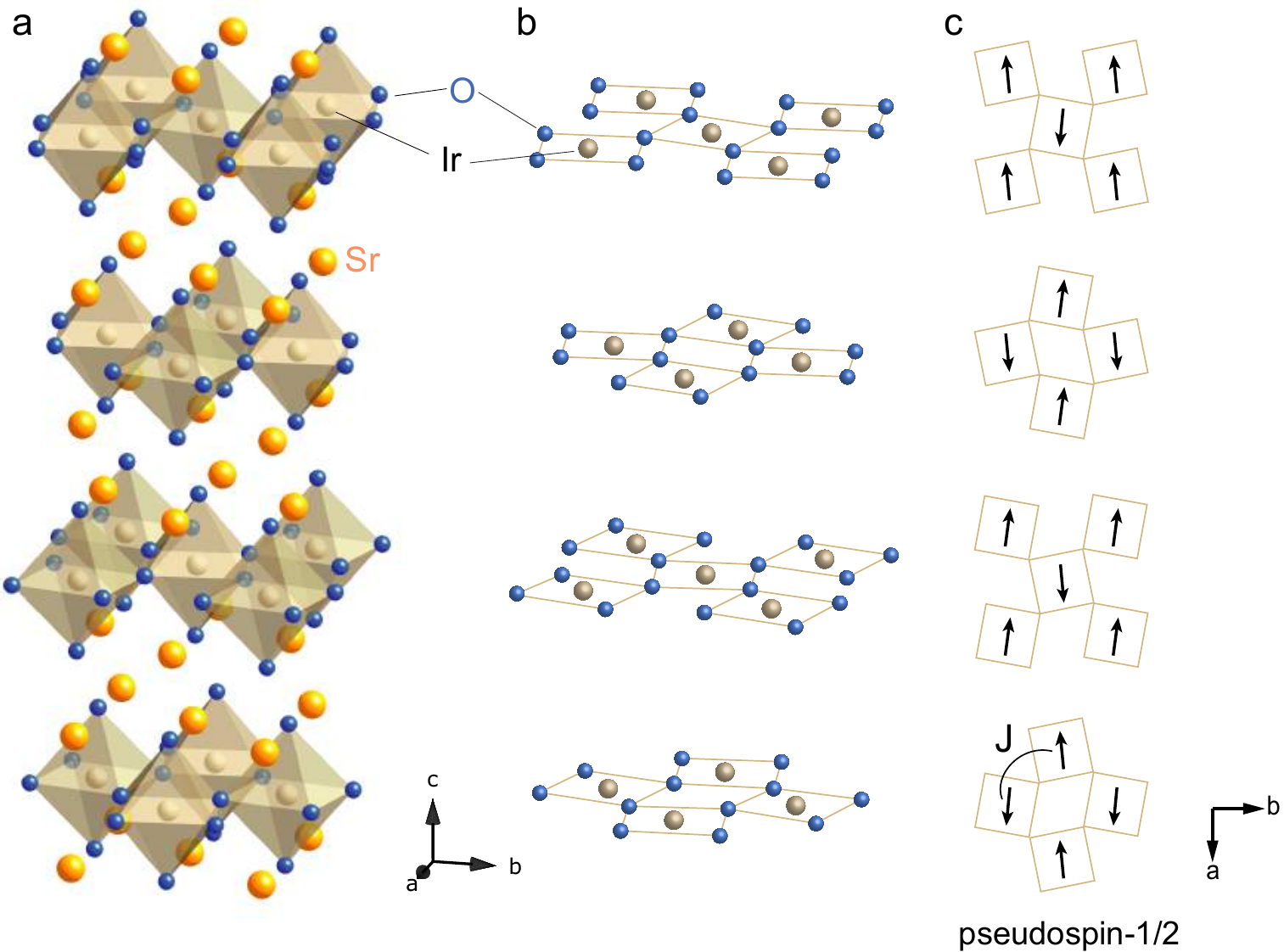}
\caption{Crystal and magnetic structures of \SIO. (a) The crystal structure is based on Ref.~\cite{Craw94} according to which the space group is I4$_1$/$acd$. However, recent studies indicate that the symmetry is lower and most probably is I4$_1$/$a$~\cite{Dhit13,Ye13a,Torc15}. (b) Quasi-2D network of Ir and O. (c) The magnetic structure from Ref.~\cite{Kim09}.}
%\label{fig-CrysMagStruc}
\end{figure}

%Section 2: Parent
\section{MAPPING ONTO CUPRATE PHYSICS}\label{sec-parent}
Soon after the discovery of HTSC in the cuprates, many complex oxides with a K$_2$NiF$_4$ structure (isostructural to La$_2$CuO$_4$) and its variants were searched for signs of superconductivity. Among these were Sr$_2$RhO$_4$ and \SIO~\cite{Subr94}, which are ``one-hole'' systems, albeit with a $t_{2g}$-hole in the low-spin $d^5$ configuration as opposed to an $e_g$-hole in cuprates. However, it was quickly apparent that $4d$ and $5d$ systems tend to be rather weakly correlated. In fact, Sr$_2$RhO$_4$ is a Fermi liquid metal~\cite{Naga10,Perr06} with a Fermi surface understood qualitatively in terms of a band structure calculated using density functional theory within the local density approximation (LDA)~\cite{Kim06,Baum06}. Interestingly, the Fermi surfaces of Sr$_2$RhO$_4$ and \SIO, calculated without SOC, are near indistinguishable due to their almost identical crystal structures with less than 1\% difference in their lattice parameters~\cite{Subr94}. A closer inspection, however, reveals that LDA does not accurately replicate the measured Fermi surface of Sr$_2$RhO$_4$, due to the presence of SOC in the $4d$ system~\cite{Liu08}. In $5d$ \SIO, SOC plays a far more dramatic role~\cite{Kim08}.

Early transport and magnetic studies noted that \SIO\ shows a significant deviation from standard Fermi liquid behavior~\cite{Craw94,Cao98}. However, the Mott insulating nature of \SIO\ went unnoticed for many years, partly due to the insufficient sample quality that obscured the intrinsic physical properties~\cite{Sung16}. Today, it is well established that \SIO\ is a canted AF insulator with a magnetic structure shown in Fig.~1. In this section, we review the mechanism of the SOC driven Mott transition, the dynamics of spin and orbital degrees of freedom governing the low-energy physics below the charge gap, and the process by which strong SOC makes it possible to map the system onto the cuprate physics.

\subsection{Spin-orbit Induced Mott Insulator}

\begin{figure}
\includegraphics[width=4in]{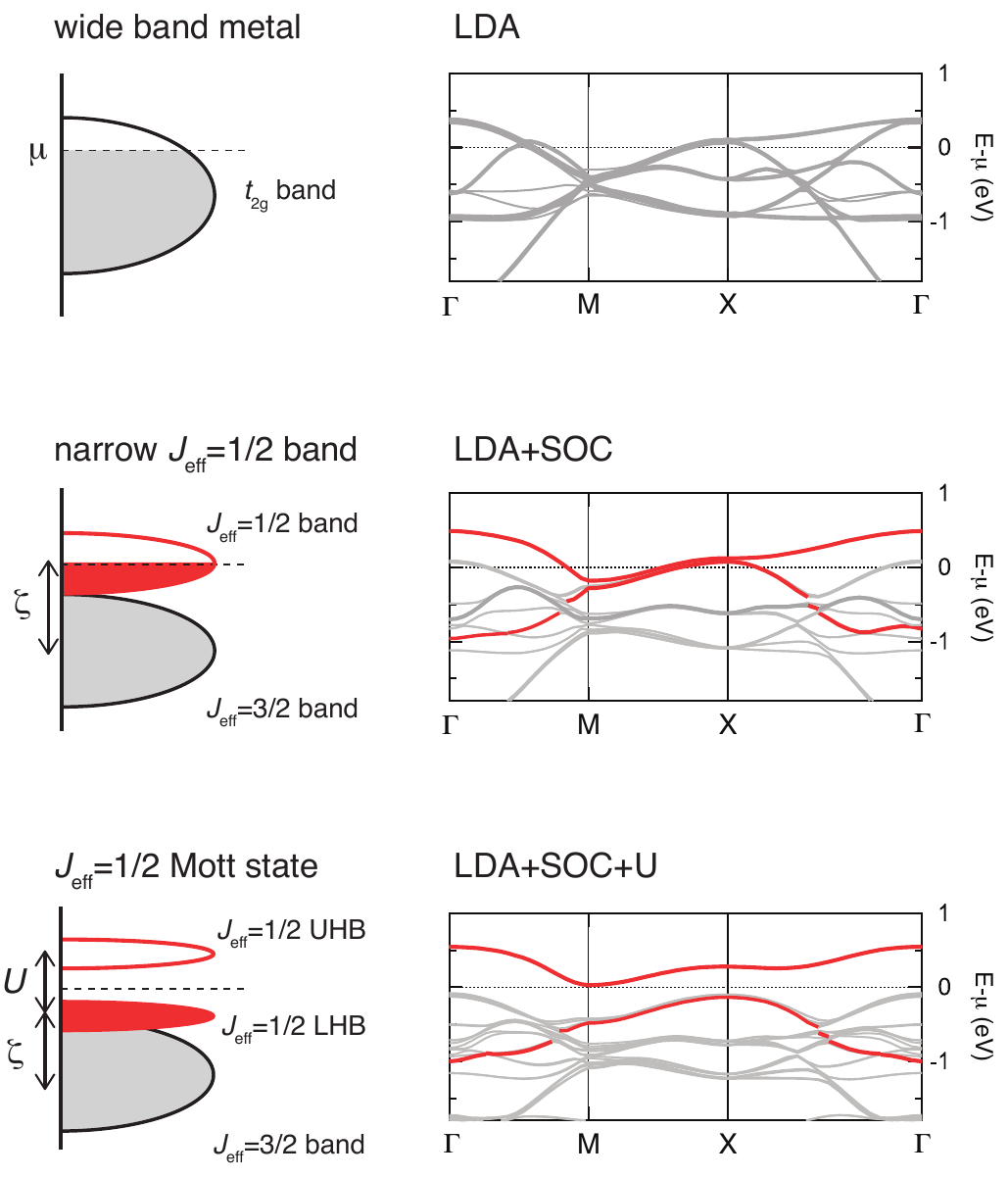}
\caption{Illustration of the SOC driven Mott transition. Introduction of SOC splits off a narrow band near the Fermi level (orange solid lines), for which a moderate Coulomb interaction $U$$\sim$2 eV is sufficient to open a gap.}
%\label{fig-SOMott}
\end{figure}

Depicted in Fig.~2 is the established picture of the SOC driven Mott transition in \SIO. In the absence of SOC, the partially filled bands of predominantly $t_{2g}$ orbital character would lead to a metallic ground state, which is precisely the case for Sr$_2$RhO$_4$~\cite{Naga10,Perr06}. Replacing the Rh$^{4+}$ ions for Ir$^{4+}$ has a very minor effect on the band structure, as long as SOC is not taken into account~\cite{Mart11}. Compared to Rh however, Ir SOC is approximately three times stronger ($\zeta$$\sim$0.4\,eV), leading to a notable modification from the metallic ground state predicted by pure LDA~\cite{Kim08}. Introducing SOC splits off a narrow band near the Fermi level, for which a moderate Coulomb interaction $U$ ($\sim$2\,eV) is sufficient to open a charge gap. Without SOC, this would require an unphysical $U$$\sim$5\,eV. Indeed, the LDA+SOC+$U$ solution captures the salient features of the insulating state of \SIO, reproducing the electronic band structure measured by ARPES~\cite{Kim08} and the canted AF magnetic structure measured by resonant x-ray scattering (RXS)~\cite{Kim09}.

From these calculations, it is clear that the insulating gap opens only as a result of the combined effects of SOC and $U$. However, there has been much debate over the correlation effects that drive the gap opening and whether \SIO\ can be classified as a genuine Mott insulator.
Although insulating above the N\'eel transition, this usual criterion for Mott insulators is not valid for 2D systems due to sizable short-range magnetic correlations that persist to much higher temperatures~\cite{Fuji12a,Vale15}.
Some have argued that \SIO\ is instead a Slater insulator~\cite{Arit12}, where the gap opens primarily as a result of magnetic order.  The observation of a significant gap reduction across T$_{\mathrm N}$, a typical Slater behavior, supports this viewpoint~\cite{Moon09,Li13}. On the other hand, the Mott picture is supported by the fact that the charge gap is an order of magnitude larger~\cite{Moon09,Nich14,Dai14,Yan15} than the magnetic energy scale $J$$\sim$60\,meV~\cite{Kim12a}. 

In the end, many theoretical and experimental studies conclude that \SIO\ has a mixed Slater and Mott character~\cite{Hsie12,Cart13,Wata14,Solo15}. In fact, there is no sharp border between a Mott and a Slater insulator and indeed it is not easy to establish any 2D magnetic system as a Mott insulator. Here instead we focus on a more pragmatic question: Are the electron correlations in \SIO\ strong enough that strong coupling theories accurately capture their physics? And vitally, do they result in the formation of exotic phenomena? In the following, we argue that many aspects of the magnetism of \SIO\ can be understood in a localized picture using formalisms developed to address the cuprate physics.

\subsection{Magnetism}\label{sec-mag}

Apart from a rather large canting angle ($\sim$12$^\circ$) that follows the staggered in-plane rotation of the IrO$_6$ octahedra~\cite{Bose13a}, the magnetic structure of \SIO\ (Fig.~1) is very similar to that of La$_2$CuO$_4$~\cite{Vakn87}. In cubic ($O_{h}$) symmetry, the ground state of the low-spin $d^5$ configuration is $J_\mathrm{eff}$=1/2 Kramers doublet, which is separated from $J_\mathrm{eff}$=3/2 quartet by 3/2$\zeta$~\cite{Abra}, where $\zeta$ is the one-electron SOC.  In tetragonal ($D_{4h}$) symmetry, the ground state remains a doublet, but their wave functions are modified through mixing with the $J_\mathrm{eff}$=3/2 quartet. In terms of $\ket{L_z,S_z}$ basis with $\ket{L_z=0}$\,=\,$\ket{xy}$ and 
$\ket{Lz=\pm 1}$\,=\,$\mp\frac{1}{\sqrt{2}}$($\ket{yz}$\,$\pm$\,$i\ket{zx})$, the doublet is expressed as 

\begin{align}
|\tilde\uparrow\rangle &=+\sin\theta\, |0,\uparrow\rangle 
-\cos\theta\, |+1,\downarrow\rangle \; \nonumber\\ 
|\tilde\downarrow\rangle &=-\sin\theta\, |0,\downarrow\rangle 
+\cos\theta\, |-1,\uparrow\rangle \;. 
\end{align}
The angle $\theta$ parametrizes the distortion through
$\tan2\theta$\,=\,$2\sqrt{2}\zeta/(\zeta+2\Delta_{\textrm{tet}})$ by 
the tetragonal crystal field splitting $\Delta_{\textrm{tet}}$. For example, $\theta$=$\frac{1}{2}\arctan 2\sqrt{2}$ in the cubic limit where $\Delta_{\textrm{tet}}$=0. 

\begin{figure}
\includegraphics[width=5in]{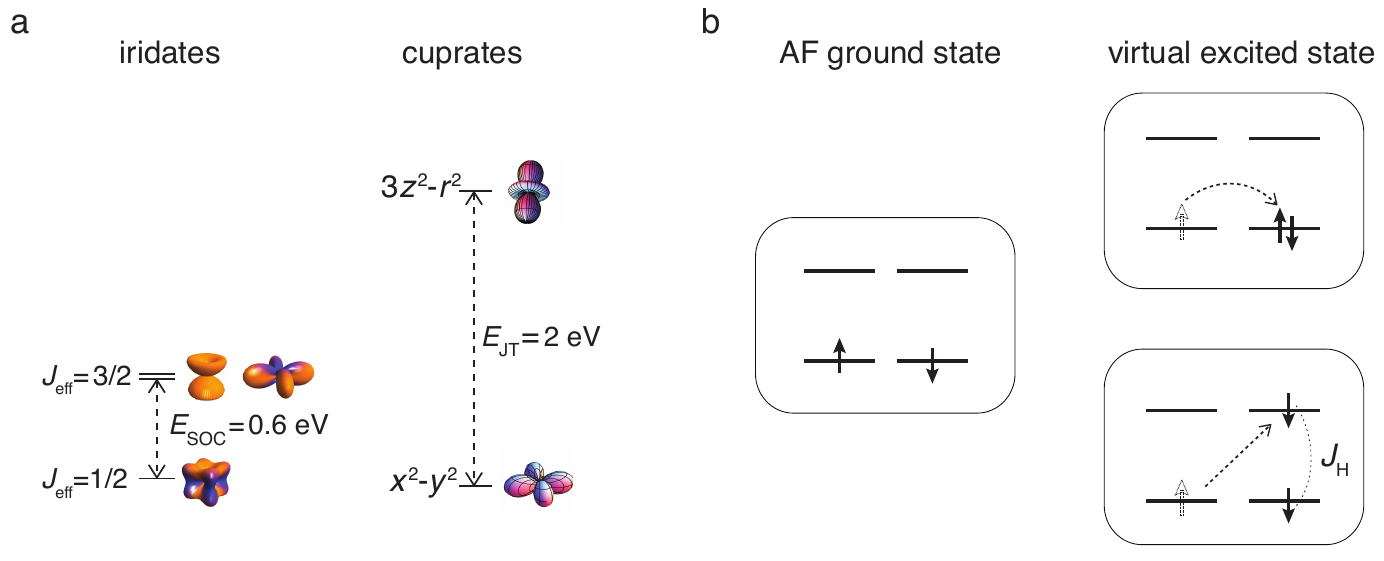}
\caption{(a) An illustration of the energy levels in iridates and cuprates. (b) Virtual charge transfer processes leading to superexchange interactions.}
%\label{fig-SpinOrbital}
\end{figure}

The energy level diagram is as shown in Fig.~3a. The doublet plays the role of spin-1/2 moments in cuprates and is referred to as `pseudospin' hereafter. It is interesting to note that the ``orbital'' degeneracy is completely lifted by the very presence of SOC, in the sense that there are no other degrees of freedom remaining apart from spin-orbit entangled pseudospin-1/2. This mechanism of orbital degeneracy lifting through SOC is very different from the Jahn-Teller effect in the cuprates, but formally they both lead to spin-1/2 dynamics in a reduced Hilbert space. The single-orbital and strongly interacting spin-1/2 moments on a quasi-2D lattice is generally regarded as an essential ingredient to HTSC and a significant effort has focused upon the search for such systems. As it turns out, this task is extremely challenging, and to the best of our knowledge \SIO\ is the only successful materialization of these physics outside of the cuprate family.

What is the nature of the magnetism expressed by the pseudospin-1/2 moments? First, we note that the orbital contribution to the magnetic moment is larger than spin~\cite{Kim08,Fuji14}, which is reflected by a minus sign in the $g$-factor. There are two conventions used in literature: one is $g$=(2,2,-2), which smoothly connects to the $g$-factor of pure $S$=1/2 moments as SOC is continuously tuned to zero~\cite{Abra}; the other is $g$=(-2,-2,-2), which has a more symmetric form and convenient in the strong SOC limit~\cite{Tho68,Chal16}. The latter convention is consistent with the wave functions defined above. Given that orbital-dominated magnetism is highly sensitive to bond geometry, one may expect a strong deviation from the usual Heisenberg interactions~\cite{Khal05}. Indeed, in a honeycomb lattice with 90$^{\circ}$ bonds isotropic interactions can be strongly suppressed, potentially leading to the Kitaev spin liquid state~\cite{Jack09}. This is not the case for a square lattice with bonds close to 180$^{\circ}$, where isotropic interactions are the leading terms. This becomes clear in the analysis of the superexchange interactions discussed in the following section.

\subsubsection{Superexchange and Magnetic Anisotropy}

%\begin{figure}
%\includegraphics[width=4 in]{SuperExchange.pdf}
%\caption{Figure caption with descriptions of parts a and b}
%\label{fig-SE}
%\end{figure}

In Mott insulators, magnetic interactions arise from virtual charge hopping back and forth between two neighboring sites. The charge transfer processes are spin independent, which means that the total spin on a given bond is conserved.
The effective spin Hamiltonian can thus be written as an inner product of the two spins; i.e. the interaction is isotropic. 
On the other hand, pseudospin generally need not be conserved during a charge transfer process because pseudospins have mixed spin-up and spin-down character. However, in a 180$^\circ$ bonding geometry the hopping arises only between orbitals of same symmetry, i.e. $xz$ to $xz$ etc., which implies that not only spin but also the orbital moment, and hence the total pseudospin, are conserved during the exchange process.
The IrO$_6$ rotations break the inversion symmetry and lead to antisymmetric exchange, but this interaction can be gauged out via a suitable rotation of the basis~\cite{Jack09,Shek92}. As a result, the single band interactions between $J_\mathrm{eff}$=1/2 pseudospins on a square lattice are entirely isotropic, just as in case of pure spins.

However, virtual excitations to $J_\mathrm{eff}$=3/2 states do allow spin flip and can lead to anisotropic interactions. This mechanism requires a nonzero Hund's coupling ($J_{\textrm H}$)~\cite{Jack09,Igar14b}, through which the pseudospin in the excited level becomes sensitive to the orientation of the pseudospin in the ground state (Fig.~3b). Thus, one arrives~\cite{Jack09} at a pseudospin Hamiltonian of the form
\begin{equation}
    H_{ij} = J\vec{S_i} \vec{S_j} + \Gamma_1 S^z_i S^z_j \pm \Gamma_2(S^x_i S^x_j - S^y_i S^y_j),
\end{equation}
where $\Gamma_{1,2}$ are functions of $\theta$ and $J_{\textrm H}/U$. Note that the last term has different signs for bonds along x and y axes and sum up to zero classical energy, and hence does not contribute to the anisotropy unless quantum fluctuations are taken into account. Thus, $\Gamma_1$ is chiefly responsible for the anisotropy, and depending on its sign it leads to either XY- or Ising-type anisotropy. Calculations show that $\Gamma_1$ changes sign as a function of $\theta$ at $\theta$=$\pi/4$, or $\Delta_{\textrm{tet}}\approx$ 190 meV~\cite{Jack09}. Thus, when $\Gamma_1$ is small, a near-ideal Heisenberg AF can be realized~\cite{Zhan13}. However, this requires a fine tuning of the relevant parameters. In fact, experiments~\cite{Kim14b,More14a} and calculations~\cite{Katu12,Bogd15,Agre17} consistently estimate $\Delta_{\textrm{tet}}$$\sim$\,-140 meV for \SIO\ and $\Delta_{\textrm{tet}}$$\sim$\,50 meV for Ba$_2$IrO$_4$\, suggesting a sizable anisotropy term $\Gamma_1$. 
%(Note the negative sign which means compressive distortion). 

\subsubsection{Overview of Experimental Results}
In parallel with the progress in our theoretical understanding of the magnetism deriving from spin-orbit entangled moments, significant advancements have been made in modern RXS techniques capable of testing the validity of the $J_\mathrm{eff}$=1/2 model.
In particular, hard x-ray inelastic RXS (RIXS) has been instrumental in the elucidation of the spin-orbital dynamics of square lattice iridates, much in the same way that inelastic neutron scattering proved invaluable to the development of the comprehensive 2D Heisenberg description of cuprates.

These studies reveal the pseudospin dynamics in \SIO\ that are remarkably similar to spin dynamics in La$_2$CuO$_4$, even as SOC is enhanced by an order of magnitude and correlations relatively weakened for the $5d$-electrons. Only the low energy dynamics around the spin wave gap mark the presence of strong SOC, indicating that the pseudospin interactions are well described by strong coupling theories. The most significant deviation from pure Heisenberg physics is an out-of-plane spin wave gap, which is a measure of the XY-type anisotropy. A smaller in-plane gap is chiefly due to pseudospin-lattice coupling, which leads to a slight structural orthorhombicity coupled to the moment orientation.

\paragraph{Probing spin-orbit entangled states}
In an elastic RXS (REXS) measurement, an incident synchrotron x-ray wavelength is tuned through the Ir $L_{2,3}$ (2$p$-5$d$) absorption edge to amplify the magnetic signal specific to the iridium ion. The resonant scattering is a second order process arising from quantum interference among all intermediate states accessible from the ground state on a given site. In this way, the REXS signal encodes information regarding the wave function for valence electrons. Calculations of the x-ray scattering matrix elements show that resonant enhancement is expected only at the $L_3$ edge for the $J_\mathrm{eff}$=1/2 state, whereas equal intensities at the $L_2$ and $L_3$ edges would arise in a $S$=1/2 system. In the first REXS study of \SIO, indeed a strong magnetic Bragg reflection signal was observed only at the $L_3$ edge, consistent with the $J_\mathrm{eff}$=1/2 model~\cite{Kim09}. However, it was pointed out that the vanishing $L_2$ intensity does not necessarily imply the $J_\mathrm{eff}$=1/2 state if the pseudospin lies in the $ab$-basal plane~\cite{Chap11,More14c,Kim17c}, which is the case for \SIO. Only the $z$ component of the pseudospin is sensitive to distortions away from the cubic limit, which means that for \SIO, measurement of dynamic fluctuations of pseudospins out of the $ab$-plane through RIXS at the $L_2$ edge can verify the $J_\mathrm{eff}$=1/2 state.

%REXS
REXS and neutron diffraction studies confirm that antiferromagnetic long range order forms below T$_{\mathrm N}$$\simeq$\,230\,K with a wavevector $k$=$(1\,1\,1)$ and pseudospin aligned in-plane~\cite{Kim09,Dhit13,Ye13a,Bose13a}. Magnetic reflections at $(1\,0\,4n+2)$ and $(0\,1\,4n)$ arise from the AF component of the magnetic structure shown in Fig.~1, while reflections at $(1\,0\,4n)$ and $(0\,1\,4n+2)$ correspond to its twin domain.  We note that the two domains have different sets of reflections because the stacking pattern is correlated with the pseudospin direction. For example, if the pseudospin are oriented along the $a$-axis, then pseudospins in (010) planes are in phase. 
The canted component is confined to the $ab$-plane, forming a net ferromagnetic moment within each layer, which orders in an $+--+$ pattern along the $c$-axis.

Above a critical field $H_C$$\approx$2\,kOe the stacking pattern rearranges in a ferromagnetic $++++$ pattern as $(1\,0\,\textrm{even})$ reflections disappear and $(1\,0\,\textrm{odd})$ emerge~\cite{Kim09}.
Although the stacking patterns along the $c$-axis is governed by small energy scale interlayer couplings~\cite{Taka16}, the complete magnetic structure vitally defines the space and point group symmetries, with consequences to selection rules in optical transitions. This is relevant to the optical second harmonic generation experiment discussed in Section~4.   

\begin{figure}
\includegraphics[width=6 in]{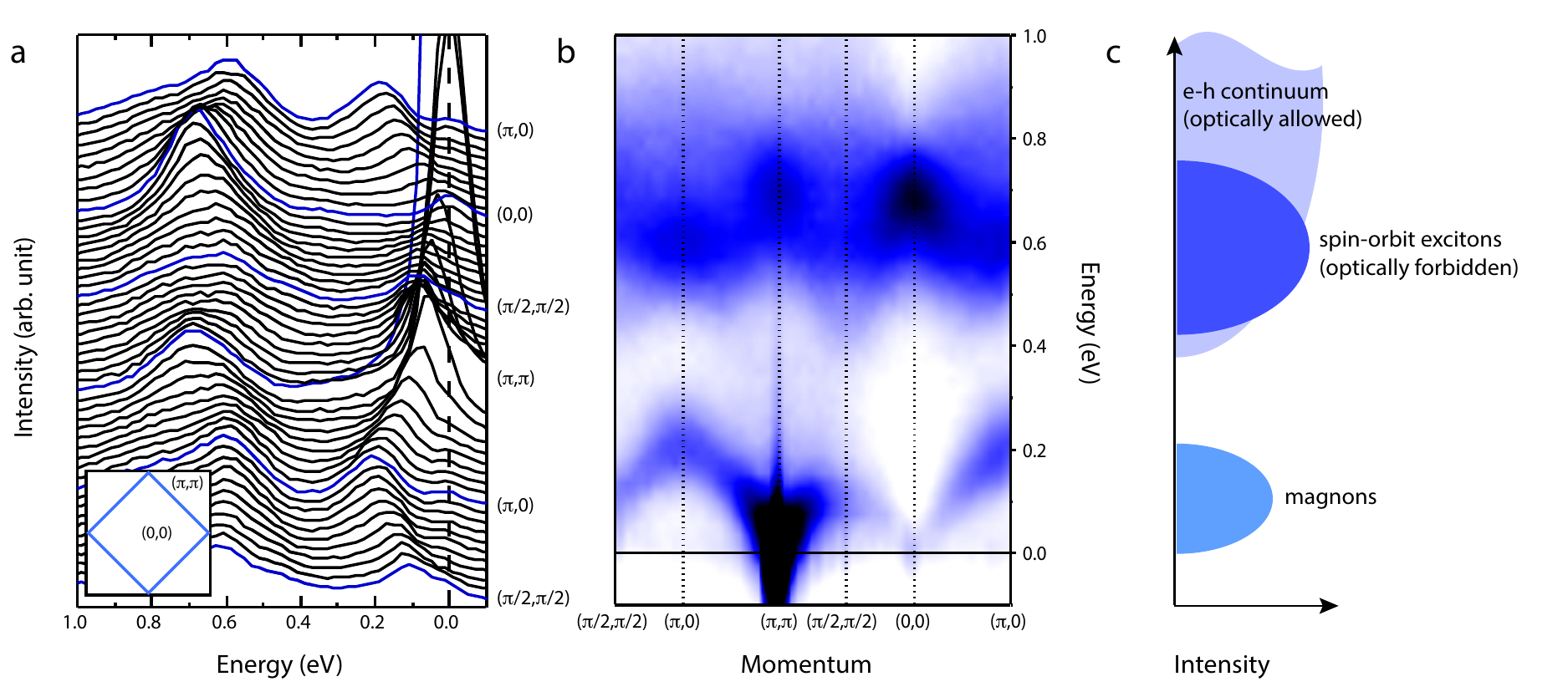}
\caption{(a,b) RIXS spectra along high-symmetry directions across the full Brillouin zone. In addition to the spin wave in the spectrum that disperses up to 200\,meV, another dispersive feature is visible at higher energies atop the electron-hole continuum. This feature is attributed to intra-ionic spin-orbit excitations across the $J_{\textrm{eff}}$=1/2 and 3/2 bands. (c) Schematic of the three representative features in the data. Figures are adopted from Ref.~\cite{Kim12a}}
\label{fig-SW}
\end{figure}

%RIXS
\paragraph{Spin waves}Direct confirmation of the predominantly isotropic nature of the exchange interactions present in \SIO\ have come from a comprehensive mapping of the full excitation spectrum using RIXS. The first published data~\cite{Kim12a}, with an energy resolution of $\sim$130\,meV, are shown in Fig.~4. Energy scans throughout the Brillouin zone reveal a single magnon mode with bandwidth of $\sim$200\,meV dispersing from the magnetic zone center ($\pi$,$\pi$). Above this spin wave branch is a distinctive feature in a range 0.4--0.8\,eV that lies on top of the electron-hole continuum formed from excitations across the Mott gap. This feature is an intra-site excitation of a hole from $J_\mathrm{eff}$=1/2 to 3/2 states (Fig.~3a)~\cite{Kim12a,Kim12d,Kim14b,Igar14a,Yang15,Lu18}. In other words, an electron-hole pair is created across the spin-orbit split bands, which form a bound state and hence is called `spin-orbit exciton'. The magnitude of the $\sim$0.6\,eV splitting of this lowest $d$-$d$ excitation can be compared to the $\sim$2 eV splitting of $e_g$ bands in the cuprates~\cite{Hozo11} (Fig.~3a); below this energy scale system can be modeled as a single band system (considering only Cu $d$ states). Similarly, excitations to the $J_\mathrm{eff}$=3/2 level are clearly separated from the single electron band and the spin wave energy scales, despite the relatively smaller splitting magnitude.

The Heisenberg-like dispersion of the spin wave is made explicit by comparing the energy and intensity of the spin wave dispersion throughout the Brillouin zone with La$_2$CuO$_4$, as shown in Fig~5. The overall similarity in both the dispersion and intensity between \SIO\ and La$_2$CuO$_4$ is remarkable. The \SIO\ magnon branch has a bandwidth of $\sim$200\,meV, as compared to $\sim$300\,meV in La$_2$CuO$_4$~\cite{Cold01,Head10}, consistent with reduced energy scales of $t$ and $U$ [see section 2.3]. Both systems are fitted with the phenomenological $J-J'-J"-J_c$ model, which takes into account respective first, second, third-neighbor isotropic exchange terms and four-spin cyclic exchange. Notably, in both systems the downturn from ($\pi$,0) to ($\pi/2$,$\pi/2$), equivalent to (1/2,0) to (3/4,1/4) in crystallographic notation, can be accounted for by either a ferromagnetic $J'$ or $J_c$, which can not be differentiated from spin wave dispersions. This dispersion is more pronounced in \SIO\ implying that the further neighbor interactions are relatively stronger as compared to La$_2$CuO$_4$, which reflects the more extended nature of 5$d$ orbitals.  To date, fitting to the highest resolution published data gives $J$=57(7)\,meV, $J'$=-18(3)\,meV and $J"$=14(2)\,meV, with $J_c$ set to zero~\cite{Vale15}.

\begin{figure}
\includegraphics[width=5 in]{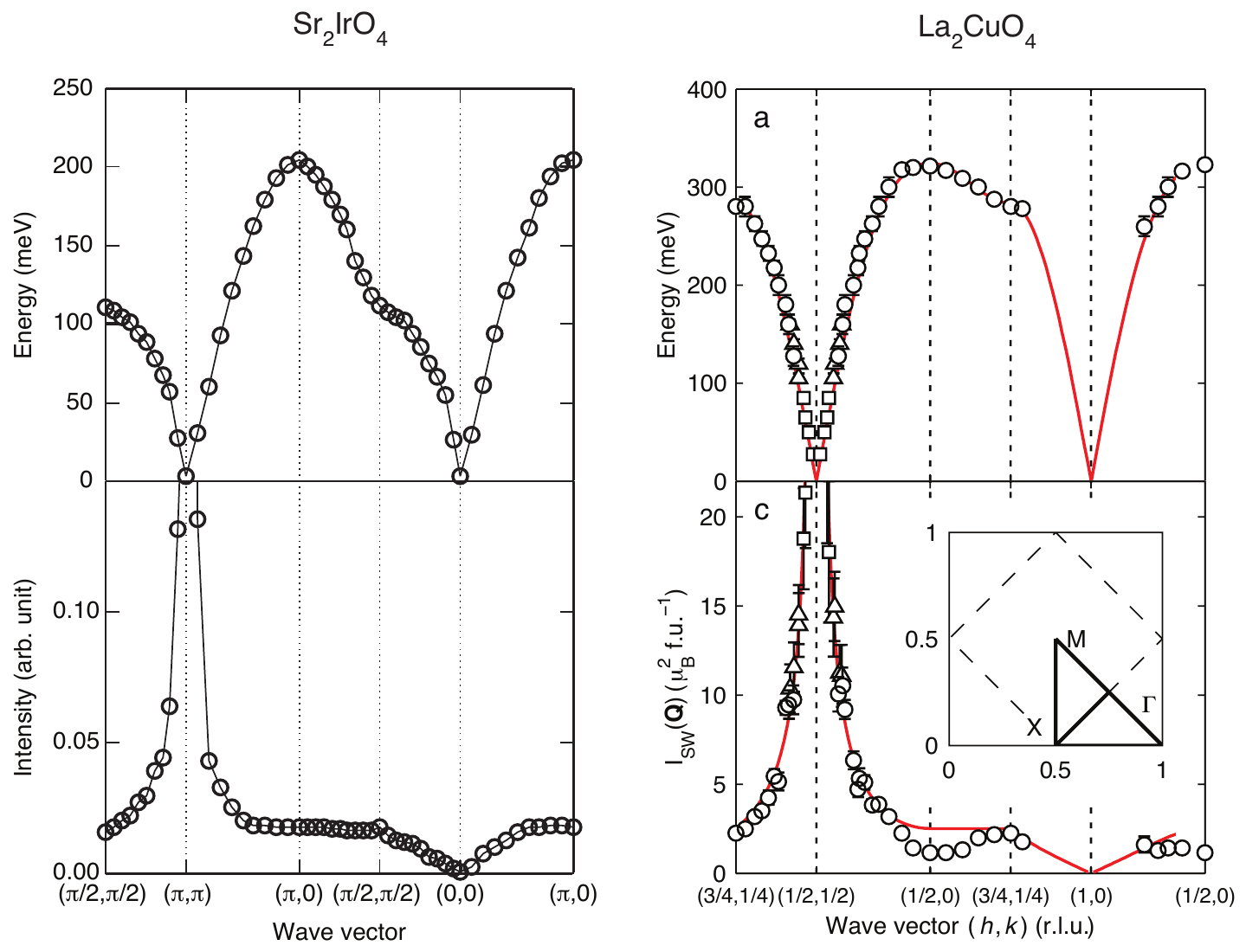}
\caption{(a)Spin wave dispersions and intensity in (a) \SIO\ measured by RIXS and (b) La$_2$CuO$_4$ measured by inelastic neutron scattering, adapted from Ref.~\cite{Head10}.}
%\label{fig-SpinWaveComp}
\end{figure}

\paragraph{Magnetic anisotropy}
A study of the low energy spin dynamics provides more insight into the magnitude and nature of anisotropic interactions, and thus the extent to which the $S$=1/2 and $J_\mathrm{eff}$=1/2 system can be compared. One approach is to probe the critical behavior by using magnetic diffuse scattering. Here, however, two separate studies present differing conclusions; one study finds critical behavior at the onset of magnetic order consistent with a Heisenberg universality class~\cite{Fuji12a}, while another indicates a significant deviation from it~\cite{Vale15}.

A more direct approach is to look at the magnitude of the spin-wave gap. RIXS measurements conducted with $\approx$50\,meV energy resolution points to a partially resolved $\approx$30\,meV spin-wave gap at the $\Gamma$ point~\cite{Kim14b,Vale15}. On the other hand, Raman scattering~\cite{Gim16} and electron spin resonance~\cite{Bahr14} measurements reveal a sharp mode at 1$\sim$2.5\,meV that develops below the N\'eel transition. Recently, this low energy feature has been resolved via high-resolution RIXS and inelastic neutron scattering~\cite{Porr17}, both of which indicate a spin-wave gap between 2 and 3\,meV at the magnetic zone center ($\pi,\pi$). While these results initially appeared incompatible, it is now clear that they correspond to different spin wave modes associated with spins rotating in the $ab$-plane and out of the plane. From the polarization dependence of the RXS process, the 30\,meV gap, measured in normal incidence, can be assigned to the out-of-plane mode. This is significantly larger than the $\approx$2.5 meV out-of-plane gap in La$_2$CuO$_4$~\cite{Pete88}. The in-plane gap of $\approx$2.5\,meV, on the other hand, is comparable to $\approx$1\,meV in La$_2$CuO$_4$~\cite{Pete88}.

Although an in-plane gap is expected from interlayer couplings and/or quantum fluctuations associated with $\Gamma_2$~\cite{Katu14}, it turns out that the in-plane gap is primarily set by pseudospin-lattice coupling~\cite{HLiu18}, which is essentially a Jahn-Teller effect. 
The magnetic ordering breaks the tetragonal symmetry of the lattice as the AF component of the pseudospins align along either $a$- or $b$ axis, which with a non-zero pseudospin-lattice coupling leads to a small orthorhombic distortion of the lattice. 
Thus, the magnitude of the in-plane spin-wave gap is a measure of the magnetoelastic coupling. Based upon the observed magnitude of the gap, an orthorhombic distortion of $\epsilon\approx10^{-4}$ is expected below $T_{\textrm N}$~\cite{HLiu18}, which is smaller than the resolution limits of most experimental techniques outside of Larmor diffraction~\cite{Nafr16} or dilatometry. There are other experimental signs, however. The pseudospin-lattice coupling is, for example, manifested as anomalous broadening of the linewidth of some Raman-active phonon modes~\cite{Gret16b,Gret17} and hardening of optical phonons across $T_{\textrm N}$~\cite{Moon09}.
The rotational behavior of pseudospins under external magnetic fields, reported in magnetometry~\cite{Taka16}, torque magnetometry~\cite{Naum17}, and magnetoresistance~\cite{Wang14} studies, is consistent with the presence of pseudospin-lattice coupling. Further, the most recent detailed analysis shows that the behavior is incompatible with four-fold symmetric anisotropy expected for tetragonal symmetry and instead is well described by a pseudospin-lattice coupling ~\cite{Porr17,HLiu18}.

The out-of-plane gap, significantly larger than that of cuprates, is a deviation from an ideal Heisenberg AF assumed in many theories of \SIO. The implications of XY anisotropy for emergent electronic orders that form through doping are currently unknown, as it is generally not found in weak SOC systems and has not been studied in any detail.
We note in passing that the bilayer compound Sr$_3$Ir$_2$O$_7$ has a strong Ising anisotropy~\cite{Kim12b,Kim12c,Fuji12b,Bose12}, which leads to pseudospin dynamics distinct from the present case and thus may have different effects on the effective interactions between doped carriers.    

\subsection{Single-band Model}
As we have seen from Fig.~3 and Eq.~1, when the symmetry is lowered from cubic symmetry the pseudospins acquire anisotropic shape through mixing between $J_\mathrm{eff}$=1/2 and $J_\mathrm{eff}$=3/2 states, which is chiefly responsible for the magnetic anisotropy~\cite{Jack09,Moha17}.
In other words, interactions among pseudospins-1/2 are renormalized by virtual excitations to $J_\mathrm{eff}$=3/2 states, but otherwise the latter disappear in the effective low-energy description of \SIO. As a result, one arrives at a single-band model~\cite{Wang11,Jin09,Wata10}, akin to the so-called $t-J$ model for cuprates, or the more extended $t-t'-t''-J$ model. 

The analogy to cuprates is made explicit by defining the pseudospins on a local basis taking into account the staggered rotation of IrO$_6$ octahedra, which gives effective hopping terms $t \approx 0.26$ eV, $t' \approx t/4$ and $t'' \approx t/10$. Note that the nearest neighbor (NN) hopping $t$ is roughly only 60\% of that in the cuprates despite the fact that 5$d$ orbitals are more spatially extended. This is because the pseudospins use orbitals of $t_{2g}$ symmetry with lobes pointing away from NN bonds, and mixing of these orbitals by SOC further reduces the overlap between NN pseudospins. However, $U$ is also smaller in \SIO\ by a similar factor, so that the ratio $t/U$ is not very different from cuprates. As a result, \SIO\ can be thought of as cuprates with the parameters scaled down to $\sim$60\%.   

One difference from the cuprates is the sign of next NN hopping $t'$, which may be responsible for the particle-hole asymmetry of the cuprate phase diagram. However, the sign can be changed by a particle-hole transformation, meaning that the hole-doped (electron-doped) side of cuprates maps onto electron-doped (hole-doped) side of \SIO~\cite{Wang11}. Indeed, as we shall see in the following sections, electron-doped \SIO\ reproduces many features of hole-doped cuprates. 

\section{ANALOGY AWAY FROM MOTT INSULATING PHASE}

\begin{figure}
\includegraphics[width=5 in]{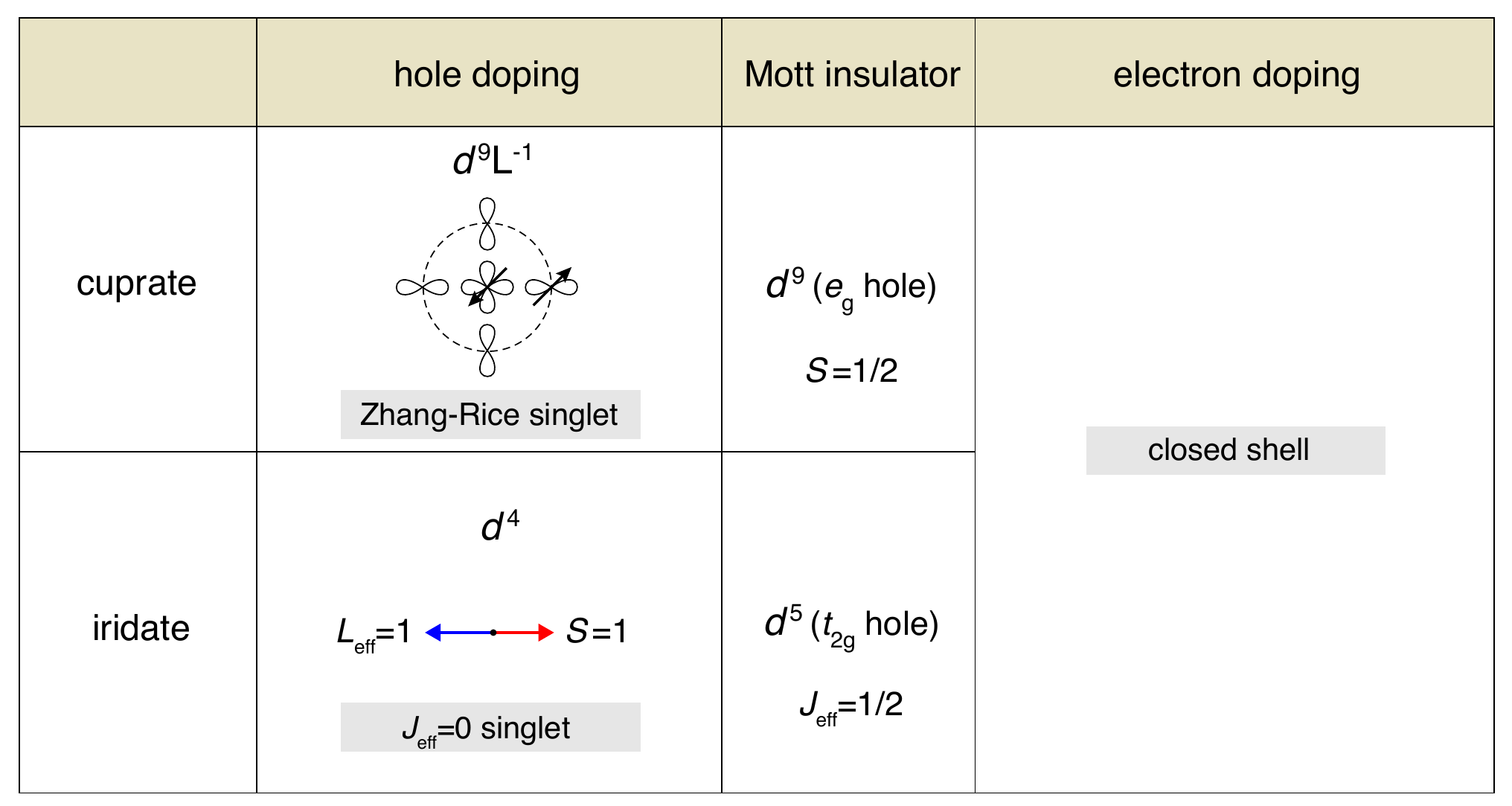}
\caption{Comparison of the structure of doped carriers in \SIO\ and cuprates.}
%\label{fig-PGE}
\end{figure}

Having discussed the analogy between \SIO\ and cuprates in their Mott insulating phases, we now proceed to compare the electronic structures in doped phases, summarized in Fig.~6. First, we note that in both cases adding an electron makes a closed shell with no internal structure. This contrasts with more complex structure on the other side with one electron removed. For cuprates, it is well known that an added hole goes into the O 2$p_{x/y}$ orbital rather than into the Cu 3$d_{x^2-y^2}$ orbital, which classifies cuprates as charge transfer insulators in the Zaanen-Sawatzky-Allen scheme~\cite{Zann85}. Thus, the ground state is a bound state of two-holes, whose two $S$=1/2 moments combine to make a so-called Zhang-Rice singlet~\cite{Zhan88}.

On the other hand, \SIO\ is a Mott insulator in the sense that the charge gap opens between $d$ bands (see Fig.~2), meaning that the doped hole goes into the $d$ band; i.e., $d^5\rightarrow d^4$ in a localized picture. Interestingly, the ground state of the low-spin $d^4$ configuration is $J_\mathrm{eff}$=\,0 singlet with antiparallel spin and orbital~\cite{Khal13}. Although this intra-site singlet formed by SOC has a different microscopic structure from that of the Zhang-Rice singlet, they share some formal similarities having identical (zero-spin) quantum numbers. 

\begin{figure}
\includegraphics[width=6 in]{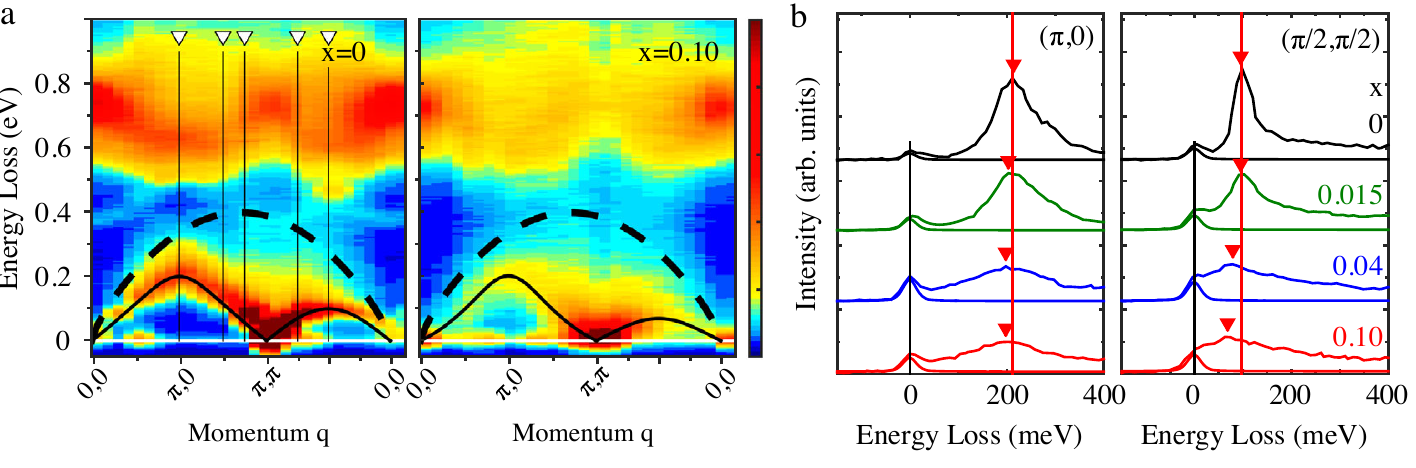}
\caption{Electron-doping evolution of spin waves. (a,b) Magnetic excitation spectra of Sr$_{2-x}$La$_x$IrO$_4$. (b) Anisotropic softening and damping of spin waves.Figures are adopted from Ref.~\cite{Gret16a}}
%\label{fig-PGE}
\end{figure}

In the cuprates, the Zhang-Rice mapping reduces the multi-orbital physics to an effective single-band physics, and its stability has been discussed in connection with the validity of the single-band model. Although the single-band model is widely accepted for cuprates, whether it contains the essential physics of HTSC continues to be debated~\cite{Varm97}. For \SIO, the validity of the single-band model is crucial for its analogy with cuprates. It is not a priori clear if and to how far the single-band picture continues to be valid away from the Mott insulating phase, because once orbital moments are quenched and SOC becomes ineffective as electrons become itinerant, the system reverts to a three-band system (as is the case for Sr$_2$RhO$_4$). This is an issue pertinent not only to the hole-doping case~\cite{Sohn16} but also to the electron-doping case~\cite{Chen16}.

In the case of electron doping, a RIXS study on La-doped \SIO, shown in Fig.~7, has revealed that paramagnon excitations persist well into the metallic phase ($\sim$$10\%$ electron doping) that has a clear Fermi surface~\cite{Gret16a}. The excitations within $J_\mathrm{eff}$=1/2 states (magnons) and to $J_\mathrm{eff}$=3/2 states (spin-orbit excitons) remain well separated and visible (Fig.~7a), indicating that pseudospin-1/2 moments persist despite rapid metallic charge fluctuations. Although heavily damped, their dispersions can be traced to reveal anisotropic softening along (0,0)-($\pi$,$\pi$) direction (Fig.~7b), which is in stark contrast to the hardening of the magnons in electron-doped cuprates~\cite{Lee14}, but resembles the results seen in holed-doped cuprates~\cite{LeT11,Dean13,Guar14}.  This validates the single-band picture for electron-doped \SIO\ and their analogy to hole-doped cuprates. 

Pseudospin dynamics in hole-doped \SIO\ has not been studied in any detail, but there are experimental indications that they map onto electron-doped cuprates. It has been shown that hole doping through replacing Rh$^{3+}$ for Ir$^{4+}$ results in suppression of the magnetic order at a much higher critical doping of $\sim$17\%~\cite{Qi12,Clan14} as compared to $\sim$3\% in La-doped \SIO~\cite{Gret16a,Chen15,Chen18}, pointing to electron-hole asymmetry as in the cuprates but with electron-hole conjugation. This is further corroborated by ARPES studies, which reveal fermiology of Rh-doped \SIO\ reminiscent of electron-doped cuprates~\cite{Cao16}.
While promising, the inference to cuprate electron-hole asymmetry is based mostly on studies of La-doped and Rh-doped \SIO, and the lack of a broader range of material families precludes making a general conclusion.

Besides chemical doping, in situ deposition of potassium atoms on a cleaved sample surface provides a clean method of electron doping~\cite{Kim14a,Kim16a}, and thus has been instrumental in studying the electronic evolution of \SIO, to be discussed in the next Section. However, this method only works at low temperatures in ultra-high vacuum and doping is limited to few surface layers, and thus can only be used in combination with surface-sensitive techniques such as ARPES and STS.

\section{EMERGENT ELECTRONIC ORDERS}
Ultimately, the analogy with cuprates is only justified by the presence of electronic orders that emerge when the magnetic order is suppressed as charge carriers are introduced--- most prominently $d$-wave superconductivity, but also phenomena surrounding the superconducting dome in the pseudogap regime. To date, the most enlightening results have come from studies of the electronic structures using ARPES~\cite{Kim14a,Kim16a} and STS~\cite{Yan15} with in situ surface electron doping. These studies reveal a pseudogap and Fermi arcs that closely follow cuprate phenomenology. Significantly, a $d$-wave gap emerges at low temperatures potentially signaling the superconducting state. In this section, we review selected experimental results that most vividly reveal such emergent phases in doped \SIO.

\subsection{Pseudogap and Competing Orders}

\begin{figure}
\includegraphics[width=6 in]{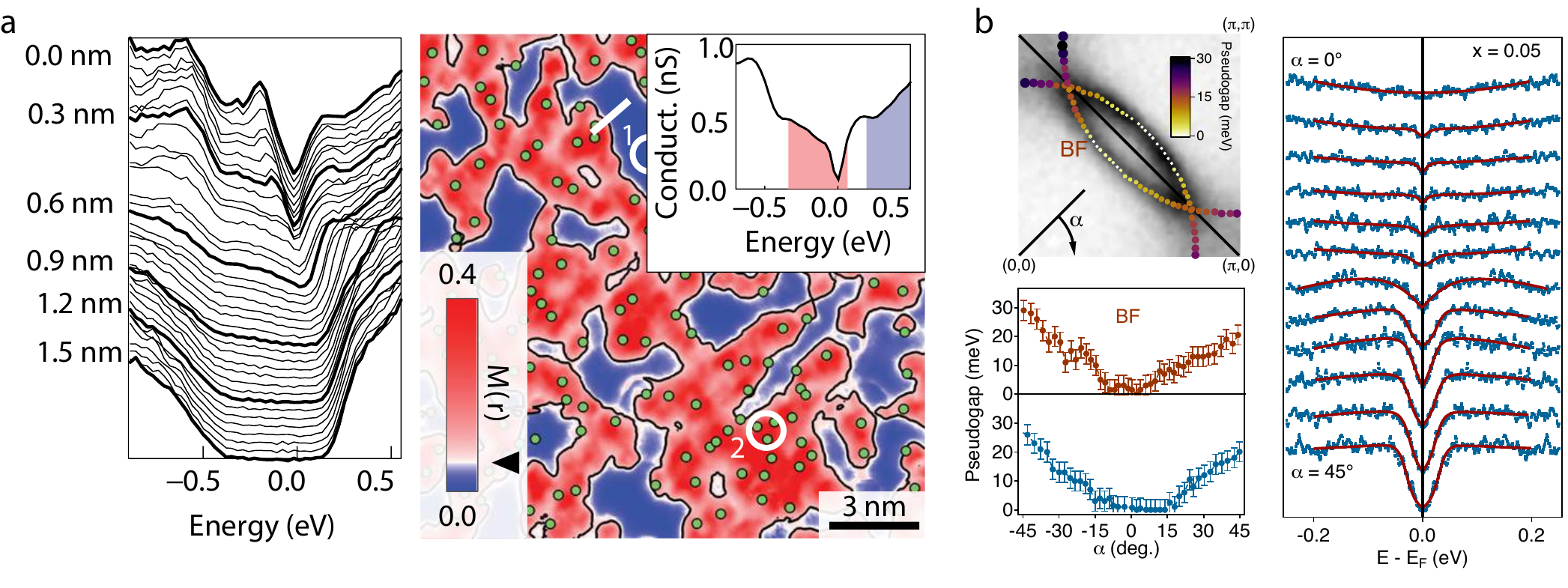}
\caption{Real and momentum space visualizations of the pseudogap. (a) (Left) STS spectra along the white line in the right panel. (Right) At 5.5\% La doping, pseudogap puddles (red) emerge separated from the pure Mott gap regions (blue). (b) A pseudogap opens along the apparent Fermi surface of 5\% La doped \SIO. Figures are adopted from Ref.~\cite{Batt17} and Ref.~\cite{Del15}.}
%\label{fig-PGE}
\end{figure}

In general, pseudogap refers to a partial suppression of spectral weight in certain regions of the Fermi surface. In the cuprate phase diagram, the pseudogap phase forms around the superconducting dome from the underdoped to slightly overdoped regions and up to relatively high temperatures. Due to its close proximity to the superconducting dome, it is widely believed that the pseudogap is key to understanding of HTSC. However, despite decades of exhaustive investigation, the origin of the pseudogap and its relation to the superconducting state remains unclear.

One thought is that the pseudogap is a precursor to superconducting gap and Fermi arcs result from broadening of the spectra that exhibit the d-wave gap~\cite{Rebe12}. In this picture, phase incoherent Cooper pairs are formed above the superconducting critical temperature ($T_c$), which is much suppressed relative to the energy scale of pair formation due to fluctuations enhanced by the reduced dimensionality. Another idea is that the pseudogap is a magnetic gap that arises due to short-range AF fluctuations~\cite{Wang15a}. A yet another possibility is that the pseudogap represents an electronic order associated with a symmetry-broken phase in competition with superconductivity. While the question remains hotly contested, it is widely believed that the underlying physics is contained in minimal models of doped spin-1/2 Mott insulator on a square lattice.

Being the first non-cuprate material to realize this condition, \SIO\ offers a unique opportunity to verify this hypothesis. We recall that \SIO\ does not have O 2$p$ band near its Fermi level and thus in this sense is a `cleaner' realization of the single-band model, especially in the electron-doped side where only the $J_\mathrm{eff}$=1/2 states are relevant. 

\subsubsection{Electron doping}
We start our discussion with electron-doped \SIO, which reproduces a remarkable range of cuprate phenomenology. In La-doped \SIO~\cite{Ge11}, as electrons are introduced a subtle change starts to be detected below 4\% doping. A RXS study shows that the long-range AF order has a slightly lower $T_{\mathrm N}$ and a reduced $c$-axis correlation~\cite{Chen18}. STS detects no change in the gap region in this doping range, suggesting that electrons are bound to dopant atoms~\cite{Batt17}. As the electron concentration is further increased, a sudden heterogeneous collapse of the Mott gap occurs and nano-scale pseudogap puddles start to appear with a gap value around 70$\sim$300 meV (Fig.~8a)~\cite{Chen15,Batt17}. In momentum space, the pseudogap opens around ($\pi$,0), leaving a broken Fermi surface referred to as `Fermi arcs' in the cuprate literature (Fig.~8b)~\cite{Del15,Brou15,Tera17}. 

Because the Fermi arcs require a radical rethinking of the notion of the topological integrity of the band, a simpler explanation has been given that they are part of the Fermi pocket that forms around ($\pi$/2,$\pi/2$) as a result of the backfolding of the bands by the ($\pi$,$\pi$) magnetic order, but only one side of it being visible due to interactions and ARPES matrix element effects~\cite{He15}. Notably, In La-doped \SIO\ there is an additional ($\pi$,$\pi$) scattering by the superstructure of IrO$_6$ rotation, which leads to an apparent Fermi pocket. This feature is not a Fermi pocket, however, as the gap is already open away from the zone boundary (Fig.~8b); i.e. there is no closed zero-energy contour. 

\begin{figure}
\includegraphics[width=4 in]{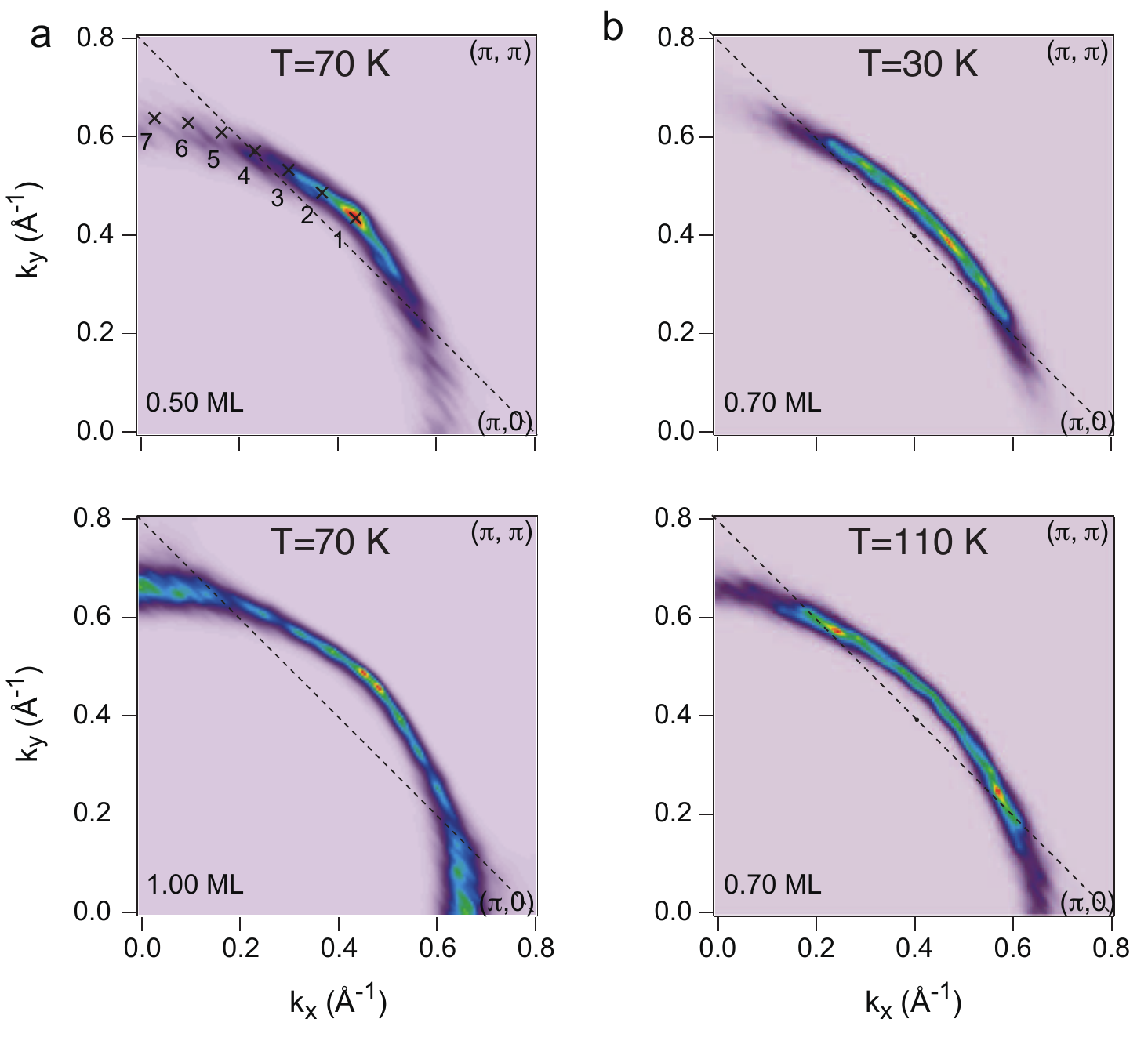}
\caption{(a) Doping and (b) temperature evolution of the Fermi arcs. Figures are adopted from Ref.~\cite{Kim14a}.}
%\label{fig-PGE}
\end{figure}

Evidence that the Fermi arcs have a nontrivial many-body origin is provided by their strong temperature and doping dependence. Using in-situ surface electron doping, achieved by depositing alkali metal atoms on a cleaved surface of \SIO, the electronic evolution over the entire range from the Mott insulating phase to the Fermi-liquid metal phase has been measured~\cite{Kim16a}. At one monolayer coverage of potassium (K), corresponding to $\sim$9\% electron doping, a circular Fermi surface encloses an area larger than one-half of the 2D Brillouin zone (Fig.~9a). However, at 0.5 monolayer, the Fermi surface breaks up into arcs as the gap opens in the region near ($\pi$,0). Furthermore, the underlying Fermi surface now encloses the an area less than half of the 2D Brillouin zone, which is inconsistent with electron doping. This implies that the pseudogap is more than merely opening of a gap as the assumption of the underlying Fermi surface violates the Luttinger sum rule. 

A similar behavior is seen as a function of temperature (Fig.~9b). At a constant doping ($\sim$0.7 monolayer), as the system is cooled from T=110\,K to T=30\,K pseudogap opening around ($\pi$,0) is clearly seen.  
Coming back to La-doped \SIO, optical conductivity below a threshold frequency of about 17 meV shows a gradual suppression of the spectral weight below $\sim$100 K~\cite{Seo17}. 

Intriguingly, a RXS study conducted in the pseudogap phase~\cite{Chen18} has uncovered an incommensurate magnetic scattering reminiscent of unidirectional spin density waves observed in the hole-doped cuprate La$_{2-x}$Sr$_x$CuO$_4$~\cite{Fuji02,Drac14}, supporting the idea that the pseudogap is associated with a symmetry-breaking order. In a similar vein, a $d$-wave spin-orbit density wave order with a circulating staggered pseudospin current was proposed to underlie the pseudogap~\cite{Zhou18}. It is claimed that the order is already present in the undoped insulating phase, manifest as the splitting of the bands at ($\pi$,0) whose two-fold degeneracy is otherwise protected by certain lattice symmetries. It is interesting to note that a magneto-electric loop-current order has been inferred from a second harmonic generation experiment~\cite{Zhao16} performed on the parent and hole-doped \SIO, although it is symmetry-wise distinct from the spin-orbit density wave. 

\subsubsection{Hole doping}

As discussed in Section~3, a doped hole has a different electronic structure than that of an electron, and is more likely to involve multi-orbital physics. This means more possibilities for emergent electronic phases~\cite{Meng14}. Known methods for hole doping of \SIO\ include K doping~\cite{Ge11} (not to be confused with K deposition discussed above) and Rh substitution in place of Ir ~\cite{Qi12,Clan14,Brou15,Ye15,Cao16,Chik17}. Although Rh and Ir are in the same column of the periodic table, their different electron affinities lead to a charge transfer from Rh to Ir, resulting in an effective hole doping of Ir sites ($d^4$) and dilution of magnetism by nonmagnetic Rh$^{+3}$ ($d^6$). Because Rh doping creates disorders in the IrO$_2$ layers~\cite{Chik15}, it would be more favorable to use K doping with dopants in SrO layers. 

Not much is known about the physical properties of K-doped \SIO. Although the authors of Ref.~\cite{Ge11} do not provide a detailed analysis, the resistivity at $\sim$4\% doping seems to show a linear temperature dependence over a wide temperature range 60$\sim$350\ K. On the other hand, Rh-doped \SIO\ 
exhibit incoherent charge transport 
characterized by the absence of quasiparticles, a momentum-independent pseudogap~\cite{Loua18}, and localization of carriers at low temperatures~\cite{Qi12}. In both K- and Rh-doped \SIO, the static magnetic order  
survives up to a much higher doping concentration ($\sim$17\%) 
as compared to electron doping.

\begin{figure}
\includegraphics[width=3 in]{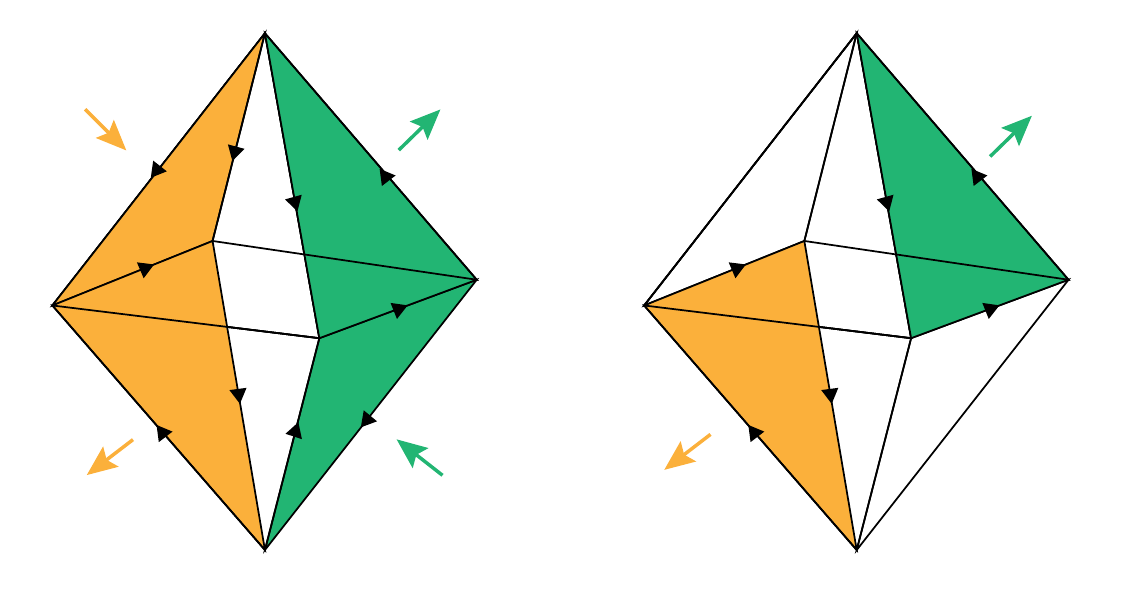}
\caption{Examples of orbital loop current orders. The arrows represent orbital magnetic moments ($m_i$) that comprise a toroidal moment defined by $\Omega =\sum_i r_i\times m_i$, where $r_i$ stand for the position and the summation is over one octahedron.
}
%\label{fig-PGE}
\end{figure}

A recent study using optical second harmonic generation on the parent and Rh-doped \SIO\ reported an electronic order that breaks the spatial inversion and rotational symmetries~\cite{Zhao16}. The technique exploits a nonlinear optical response of a material and is highly sensitive to changes in the point group symmetry. The onset of the signal nearly coincides with the N\'eel temperature in the parent compound, but the two temperatures diverge with increasing Rh-doping as the AF transition is more quickly suppressed. A symmetry analysis shows that the data is consistent with an orbital loop-current order, which was proposed to underlie the pseudogap in cuprates~\cite{Varm97,Fauq06}. Assuming that the time-reversal symmetry is also broken, magnetic space groups consistent with the data are narrowed down to $2'/m$ or $m1'$, for which different loop-current patterns can be constructed (Fig.~10). Subsequently, a polarized neutron diffraction measurement found a peak in a spin-flip scattering channel whose onset temperature matches that of the second harmonic generation, supporting the orbital loop-current interpretation of the second harmonic generation experiment~\cite{Jeon17}. 

Loop current-ordered phases, however, are based on a three-band model including O 2$p$ states---a feature not present in a hole- or electron-doped \SIO. There are, in fact, simpler explanations for the second harmonic signal~\cite{DiM16}. The $2'/m$ magnetic space group is also consistent with the $-+-+$ stacking pattern that can readily form with minor perturbation through disorder, doping, or laser pump. It has been pointed out that the inversion symmetry can also be broken locally at a boundary between two magnetic domains through magneto-elastic coupling~\cite{HLiu18}, and particularly in the paramagnetic phase with short-range magnetic fluctuations the lattice can appear inversion symmetry broken depending on the timescale of the measurement.

\subsection{Superconductivity}
\begin{figure}
\includegraphics[width=6 in]{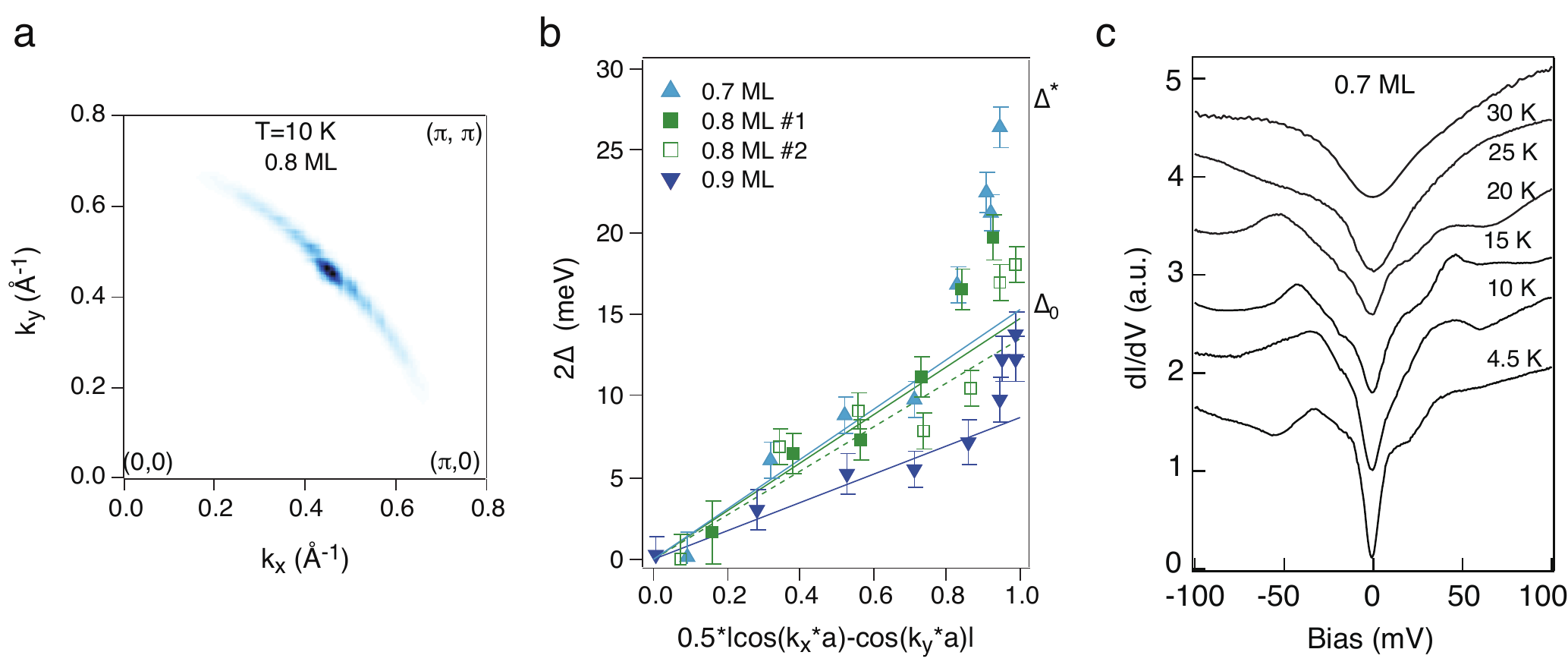}
\caption{Possible superconducting gap in in situ electron-doped \SIO. (a) Fermi arc shrinks to a point node upon cooling below T$\sim$30\,K as a $d$-wave symmetric gap opens along the arc. (b) Gap magnitude as a function of $d$-wave form factor. (c) Temperature dependence of the V-shaped gap in STS spectra. Figures are adopted from Ref.~\cite{Kim16a} and Ref.~\cite{Yan15}.}
\label{fig-dwave}
\end{figure}

Although there is no experimental evidence of superconductivity in electron or hole-doped \SIO\ at the time of writing, tantalizing indications of $d$-wave superconductivity have been glimpsed in in situ electron-doped \SIO. An ARPES measurement shows that below $\sim$30\,K the Fermi arcs shrink to point nodes as a $d$-wave-symmetric gap opens along the arc (Fig.~11a)~\cite{Kim16a}. A fit to the $d$-wave form factor, however, reveals a prominent deviation near the antinode (Fig.~11b).  A linear extrapolation of the near-nodal gap returns a gap maximum in the range 16-30\,meV, depending on the doping concentration (or the surface potassium coverage). This is smaller by a factor of two than the actual gaps $2\Delta^*$ measured at the anti-node, pointing to the possibility that the pseudogap and the $d$-wave gap might have two distinct origins. In fact, this is a feature observed universally in underdoped cuprates~\cite{Yosh12,Hash14}, and thus \SIO\ can shed new light onto the elusive relationship between the pseudogap and the $d$-wave gap. 

In real space, the pseudogap and the $d$-wave gap show anticorrelations~\cite{Yan15}, again in much similarity with the STS measurements on the cuprates. The $d$I/$d$V spectra exhibit a strong spatial inhomogeneity with high-energy-gap regions separated from low-energy-gap regions. The low-energy gap opens symmetrically about the zero bias energy below a doping-dependent temperature, ranging between 20$\sim$50 K for the surface coverage between 0.5$\sim$0.7 monolayers, with the gap magnitude in the range 10$\sim$28 meV. We note that the results for 0.7 monolayer (which is the only overlapping data point) are in good agreement with the ARPES results (Fig.~11c).  

Based on the above results it is tempting to associate the $d$-wave gap with superconductivity. Although theoretically $d$-wave gap is not exclusive to superconductivity~\cite{Mars89,Chak01}, no other phase of matter has been observed to exhibit a $d$-wave gap symmetric about the Fermi level. However, establishing superconductivity requires macroscopic evidence, i.e., Meissner effect or zero-resistance, which are very difficult to measure on in situ doped samples. We note that La-doped samples only partially reproduce these results~\cite{Del15}; high-temperature pseudogaps are observed but no there are no $d$-wave gaps, but this may only be due to insufficient sample quality~\cite{Sung16}.

\section{SUMMARY AND OUTLOOK}

Over the last decade, the field of iridates has grown to a size that a single review article can no longer cover all of the exciting physics they encompass. In this review, we focused on one material system, the square lattice iridate \SIO, for which the central issue has been whether it would display HTSC like cuprates. Even staying within this somewhat narrow scope, limited space did not allow us to touch upon different approaches to metallize \SIO\ through epitaxial thin films~\cite{Serr13,Nich13}, high pressure~\cite{Hask12,Zocc14}, gating~\cite{Lu15,Ravi16}, and other chemical routes~\cite{Okab13}. So far, the most promising results have been obtained from in situ doping used in combination with ARPES and STS, but the extreme surface sensitivity of these methods is a serious impediment to further scientific progress.

What is the physical origin of the low temperature $d$-wave gap observed in in situ electron-doped \SIO, and why is it not reproduced in La-doped \SIO?  These are important outstanding questions, because if the $d$-wave gap is not due to superconductivity, it at the very least represents a new phase of matter likely in competition with $d$-wave superconductivity. 
One hint is that the $d$-wave gap phase may be very fragile against disorder. It is important to note that even the parent compound \SIO---much more so for the La-doped samples---is difficult to grow in high quality as it is prone to contamination by oxygen vacancies~\cite{Guer18}, which may mask the intrinsic properties of parent~\cite{Sung16} and possibly destroy the coherence of $d$-wave gap in electron-doped \SIO. In fact, a well-defined quasiparticle is only present in the ARPES spectra of in situ doped \SIO\ but not in La-doped \SIO. In the former, close to one monolayer coverage the system is almost disorder free, and the off-plane charge reservoir also acts toward screening any charged impurities near the sample surface where the $d$-wave gap is observed. The destructive effect of poorly screened 
Coulomb impurities on electronic quasiparticles is manifested in the insulating \SIO\ by charge-neutral (spin-orbit) excitons, which, despite their much higher excitation energies, are much more long-lived than fermionic quasiparticles~\cite{Kim14b}. 

These results suggest that methods beyond traditional chemical doping, such as digital doping through heterostructures or hydrogenation may be the way to proceed. The relatively simpler chemistry of the cuprates, also the pnictides for that matter, fueled the explosive growth of the HTSC field. 
But the chemistry and physics differences between iridates and cuprates may be used to our advantage in solving the HTSC conundrum, by separating material specifics and issues idiosyncratic to cuprates from essentials of HTSC. 
In particular, although both iridates and cuprates are single-band systems at low energies, they are derived from Mott and charge-transfer insulating parent compounds, respectively.   
The results from electron-doped \SIO\ already tell us that this distinction seems not to be essential for 
the emergence of pseudogap and formation of $d$-wave gapped Fermi surface, and that these phenomena can be captured by a single-band model interacting with short-range AF fluctuations. Future studies of square-lattice iridates may provide new insights on whether Mott versus charge-transfer dichotomy is essential for HTSC itself. 

In solving these problems, one can use material design and experimental probes not applicable to cuprates. For example, the fact that the bilayer compound Sr$_3$Ir$_2$O$_7$ with strong Ising anisotropy does not show any sign of unconventional charge dynamics when doped~\cite{Del14,Hoga15} suggest that a clever engineering of spin dynamics can lead to novel tunable material properties. Such materials design can be readily tested on a small single crystal using hard x-ray RIXS with its access to spin dynamics over the full Brillouin zone up to very high energies with energy resolutions now getting better than 10 meV~\cite{Kim18}. Although not discussed in this review, iridates are also one of the major playgrounds for spin liquids~\cite{Okam07,Kita18}, which has a historical connection to HTSC~\cite{Ande73}. Beyond a cuprate analog, iridates are thus interesting in their own right and at the moment we are probably seeing only the tip of the iceberg.

%Disclosure
%\section*{DISCLOSURE STATEMENT}
%The authors are not aware of any affiliations, memberships, funding, or financial holdings that
%might be perceived as affecting the objectivity of this review. 

% Acknowledgements
%\section*{ACKNOWLEDGMENTS}
We acknowledge insightful discussions with B.~Keimer, G.~Jackeli, H.~Takagi, H.~Gretarsson, M.~Le Tacon, J.~van den Brink, J.~H.Kim, Y.-B.~Kim, H.-Y.~Kee, C.~Kim, J.-H.Park, Y.~Bang, and E.~G.~Moon.

% References
%\bibliographystyle{ar-style4}
\bibliography{sio214rev}

\end{document}